# Optimization of microfluidic synthesis of silver nanoparticles: a generic approach using machine learning

*Konstantia Nathanael\* [1], Sibo Cheng [2], Nina M. Kovalchuk [1], Rossella Arcucci [2,3] and Mark J.H. Simmons [1]*

[1]School of Chemical Engineering, University of Birmingham, UK

[2]Data Science Institute, Imperial College London, London, SW7 2AZ, UK

[3]Earth Science & Engineering Department, Imperial College London, London, SW7 2AZ, UK

*\*Corresponding Author's email:* cxn782@student.bham.ac.uk

**Abstract**

The properties of silver nanoparticles (AgNPs) are affected by various parameters, making optimisation of their synthesis a laborious task. This optimisation is facilitated in this work by concurrent use of a T-junction microfluidic system and machine learning approach. The AgNPs are synthesized by reducing silver nitrate with tannic acid in the presence of trisodium citrate, which has a dual role in the reaction as reducing and stabilizing agent. The study uses a decision tree-guided design of experiment method for the size of AgNPs. The developed approach uses kinetic nucleation and growth constants derived from an independent set of experiments to account for chemistry of synthesis, the Reynolds number and the ratio of Dean number to Reynolds number to reveal effect of hydrodynamics and mixing within device and storage temperature to account for particle stability after collection. The obtained model was used to define a parameter space for additional experiments carried out to improve the model further. The numerical results illustrate that well-designed experiments can contribute more effectively to the development of different machine learning models (decision tree, random forest and XGBoost) as opposed to randomly added experiments.

**Highlights:**

- Decision tree guided synthesis is an effective strategy to design experiments.
- Three models were used to test the performance of each area of parametric space.
- The proposed method allowed the use of kinetics to predict the size of AgNPs.

**Abbreviations:**

AgNPs – silver nanoparticles;

AI – artificial intelligence;

BO – Bayesian optimization;

DCA – dicarboxyacetone;

De – Dean number;

DNN – deep neural network;

DT – Decision Tree;

MAE – Mean Absolute Error;

ML – machine learning;

MSE – Mean squared error;

PTFE – polytetrafluoroethylene;

PVA – polyvinyl alcohol;

Re – Reynolds number;

RF – Random Forest;

RRSE – Root Relative Square Error;

SI – supporting information;

SN – silver nitrate;

TA – tannic acid;

TC – trisodium citrate;

XGBoost – Extreme Gradient Boosting.

# 1. Introduction

The precise and controlled fabrication of silver nanoparticles (AgNPs) is required for a range of applications [1]. Optimisation of the chemical synthesis process is a very labour-intensive, costly and time-consuming task because it includes multiple reagents and different independent experimental conditions including the types and concentrations of reactants, temperature, reactor design and mixing conditions. For example, the size of AgNPs strongly depends on both the type and concentration of stabilizing agent [2],[3], [4] and often depends on pH of the solution [5],[6],[7]. Ionic stabilizers such as citrate used in this study produce a charged layer around particles which inhibit agglomeration. Several studies showed that both decreasing [8] and increasing [9] particles sizes can be obtained by varying the amount of stabilizer. Das, Bandyopadhyay [10] and Henglein and Giersig [11] found there is an optimum concentration of trisodium citrate in AgNP synthesis. They observed that smaller particles with a narrow size distribution can be formed for a specific range of citrate concentrations and larger particles at concentrations higher than the optimal, due to the high ionic strength of the solutions. It was observed that the reducing ability of the completely hydrolysed citrate species and dicarboxyacetone (a by-product of citrate reaction with silver) is higher at pH 12, causing a fast reduction rate of the precursor and therefore smaller particles [12]. However, it was also found in Liu, Kozlovskaya [13] that alkaline pH in the presence of high salt concentrations can lead to thicker tannic acid-poly(N-vinylpyrrolidone) (TA-PVPON) multilayers as a result of screening of the negatively charged tannic acid. Furthermore, the hydrodynamic conditions play a significant role in the nucleation and growth steps for the synthesis of nanoparticles [14],[15]. Relevant works are represented in Khan, Günther [16] and Wu, De Varine Bohan [17]. The former demonstrated that the sizes and size distributions of colloidal silica particles can be tuned in laminar flow reactors and segmented flow reactors by varying their linear flow velocity and mean residence time. The latter showed that decreasing the helix diameter increases the mixing in helical reactor (due to generation of Dean vortices) which then leads to a controlled size distribution in continuous synthesis of AgNPs in the absence of capping agents. A similar effect was observed at increasing flow rates when the Dean number ($De$) was above 5.

Microfluidic technology can improve the optimisation process by providing better control over the reactions and reduced reagent consumption [18]. Machine learning combined with microfluidic synthesis offers a promising approach to tackle the repeated and extensive experimental tests and accelerate the development of efficient protocols [19].

Recently, artificial intelligence (AI) has become a valuable aid for many scientific and engineering fields, including nanotechnology, due to its contribution in data acquisition and processing improvements [20], [21],[22]. Machine learning (ML) algorithms which are a subset of AI have been employed specifically in the synthesis of AgNPs to predict their characteristics [23],[24],[25] or perform other kinds of decision making under uncertainty [26]. Mekki-Berrada, Ren [27] combined a deep neural network (DNN) and Gaussian process-based Bayesian optimization (BO) to synthesize AgNPs with a desired absorbance spectrum in a droplet microfluidic device. They used the flow rates of silver seeds ($Q_{seed}$), of silver nitrate, SN ($Q_{AgNO_3}$), of trisodium citrate, TC ($Q_{TC}$) and polyvinyl alcohol, PVA ($Q_{PVA}$) as input parameters and efficiently predicted the desired plasmon resonance for the reaction synthesis. Sattari and Khayati [28] applied a Gene Expression Programming (GEP) to predict the size of AgNPs prepared by a green synthesis route. The proposed predictive model with coefficient of determination, $R^2$ = 0.9961, mean absolute error, MAE = 0.2545, and root relative square error, RSME= 0.0668 showed that the initial concentration of silver precursor and plant extract were the most influential parameters for the final particle size. Other relevant work used to predict the final particle size of AgNPs is shown in Shabanzadeh, Senu [29].

Decision trees are an established keystone in machine learning literature [30] as a powerful tool for prediction, interpretation and data manipulation. More specifically, a decision tree is a supervised learning algorithm which sorts a population into segments. This approach looks like an inverted tree with a root node, some internal nodes and leaf nodes. The root node is always on the top of the tree structure and represents the variable that best splits the data, whilst the leaf nodes at the bottom display the final outcome of a combination of decisions. A decision pathway is presented by a line connecting the root node with one of the leaf nodes [31]. Compared with deep learning approaches [32], the advantages of a tree-based algorithm are the good interpretability of the constructed models and the low computational cost. It can identify, in a linear way, information and relationships that can be used to design future experiments and analyse data [30]. A limitation is that it can display overfitting or underfitting within a small data set [31].

The literature review revealed that previous ML studies for the synthesis of AgNPs focused on input parameters which are related to specific chemical compositions and process conditions. This means that a model which is derived for a certain experimental configuration cannot be applied for another set of reagents or for a different reactor design. The aim of this study is to develop a generalised workflow to design experiments for fast prediction of the size of AgNPs synthesized in microfluidic systems. It is suggested that the chemistry of the process can be described in the terms of nucleation and growth constants, whereas the hydrodynamics of the process can be accounted for by Reynolds number, *Re,* and the ratio of Dean number to Reynolds number $De/Re$ . Storage temperature is included to enable consideration of particle stability after collection.

A machine learning-guided design of experiments based on the decision tree method has been applied for the production of AgNPs in a continuous flow microreactor. The proposed strategy, which used optimal and flexible experiment designs based on uncertainty analysis, can provide more flexibility compared to existing approaches such as design of experiments [33], since the exact values of the initial parameters are not fixed and the training features could be parameters which have been calculated based on physical parameters, such as nucleation and growth constants. The nucleation and growth constants were derived from an independent set of experiments carried out in a beaker using the Finke-Watzky (F-W) two-step mechanism [34] which has been applied previously to describe well a wide range of kinetic processes such as protein aggregation [35],[36] and metal nanoparticle formation including silver [37], gold [38], palladium [39] and rhodium [40]. Only one set of chemicals was used in this study providing several pairs of nucleation and growth constants. These kinetic constants were used as input parameters in the decision tree to build a ML model which combines parameters that can be applied in the future for different sets of chemicals. By applying this strategy, a general model for the synthesis of AgNPs can be obtained in a more direct way by providing the region of interest for further experiments. The suggested approach can also enable a considerable reduction of ecological impact of process optimization by using microfluidics, which manipulates small samples under well-controlled conditions resulting in a considerable reduction of materials and energy consumption.

## 2. Materials and methods

### 2.1 AgNPs synthesis and characterisation

The silver nanoparticles were synthesized in a T-junction microfluidic device composed of 0.5 mm inner diameter polytetrafluoroethylene (PTFE) cylindrical tubes (Cole-Parmer). Two inlets (length of 0.6 m each) were used to supply silver nitrate, SN (0.92 mM) and the mixture of tannic acid, TA (0.123 mM) / trisodium citrate, TC (1.91 mM - 3.82 mM.) respectively and an outlet tube of 2 m length was used as the reaction channel (Fig. 1). The inlet flow rates were regulated by syringe pumps (World precision instrument-AL-4000) equipped with 5 mL syringes (Fisher), while the three tubes were connected with a Tee tubing junction (0.020" (0.5 mm) Thru-Hole) from Upchurch Scientific. The outlet channel has the same cross section as the inlet channels, therefore the superficial velocity in the in the outlet channel is twice that observed in each of the two inlet channels.

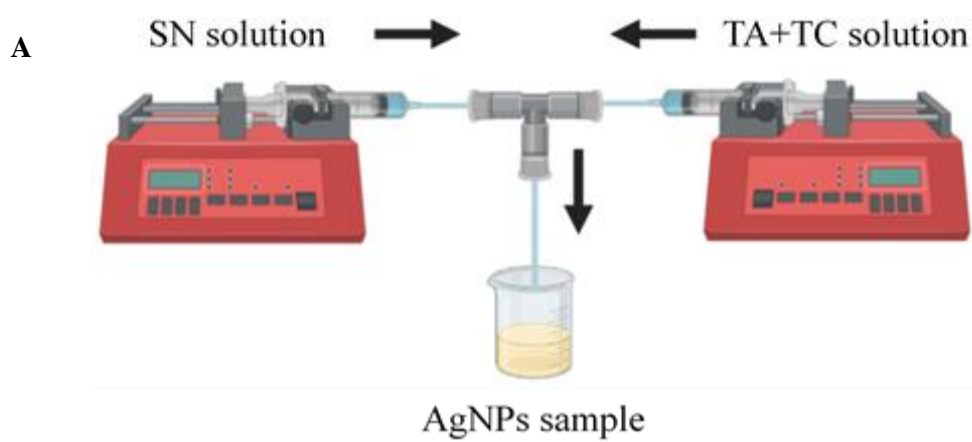

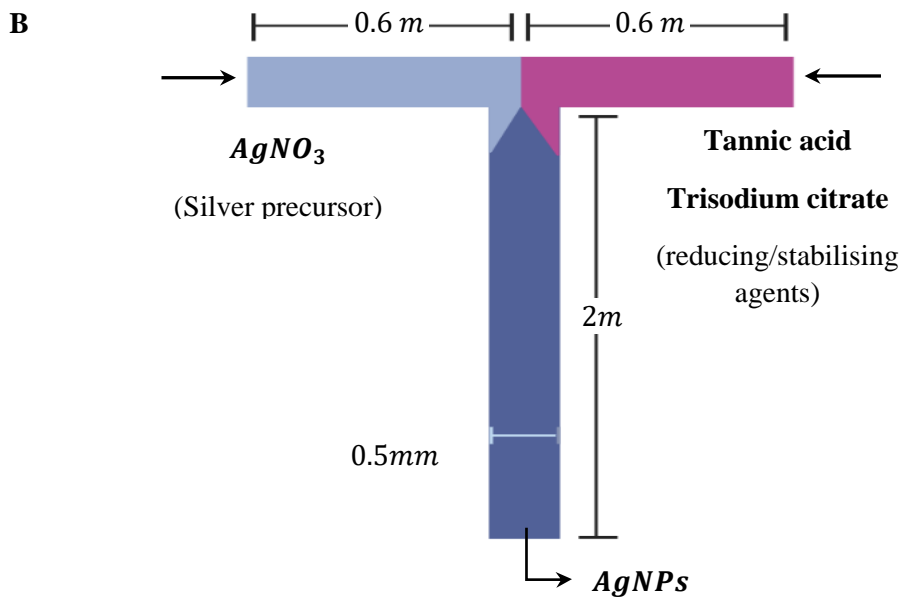

Figure 1: A) Experimental set up B) Schematic diagram of T-junction microfluidic device.

The concentrations of reagents were determined based on a previous study of Kašpar, Koyuncu [41] with modifications in the concentrations of reducing agent (TA) and stabilizing agent (TC). The solution of TA and TC had a pH of 7 or 12. The reagents including silver nitrate, 99+% ($AgNO_3$), tannic acid ($C_7H_{52}O_{46}$), trisodium citrate dihydrate ($Na_3C_6H_5O_7 \cdot 2H_2O$) and sodium hydroxide ($NaOH$) were purchased from Alfa Aesar.

The outlet tube was either straight or coiled onto a 3D-printed helical shape devices with diameters of 3 mm and 5 mm to improve the mixing of the reagents. The particle size was measured using dynamic light scattering (DLS) by Zetasizer Nano series (Malvern) and confirmed by transmission electron microscopy (TEM) (JEOL JEM-1400). The concentration of silver nitrate at different flow rates was measured to investigate the completeness of the reaction. A silver electrode from EDT direction connected to mV meter from Mettler Toledo-FP20 was used.

The nucleation, $k_1$, and growth, $k_2$, rate constants were derived from the absorbance intensity at 400 nm, characteristic for AgNPs using F-W two step mechanism [34] which combines a homogenous nucleation reaction and an autocatalytic growth process in which nuclei and growing particles play the role of auto-catalysts. The time dependence of the absorbance of AgNPs (kinetic curve) for different experimental conditions was measured by UV-vis spectrophotometer (Jenway-6300). Reactions were carried out in a beaker and mixing was set at a point where further increase of mixing intensity does not affect the kinetic curve. Thus, AgNPs synthesis was independent of mass-transfer and the rate was determined by true chemical kinetics. To quantify the amount of silver in AgNPs at any time during synthesis, the UV-vis spectra was recorded every 5 min. All measurements were performed in triplicate.

The synthesis of AgNPs involves three different steps [42] including the reduction of silver ions to silver atoms, the formation of silver nuclei and their subsequent growth to generate AgNPs (see Fig. 2). In this study, the reaction was performed through the reduction of SN in the presence of TA which is a weak reducing agent and TC which is both reducing and stabilizing agent.

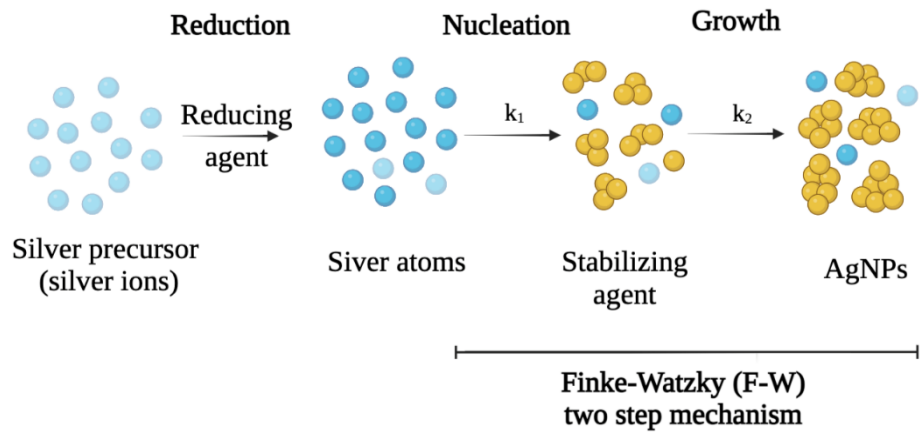

Figure 2: Formation of AgNPs through three different steps, i) the reduction of silver ions to silver atoms, ii) the nucleation step where the smallest thermodynamically stable clusters are formed and iii) their growth to produce AgNPs. The nucleation and growth step can be described based on F-W mechanism, where reduction rate is included in nucleation rate.

As described in Eq. (1) and (2), nuclei represented as $B$ are produced uniformly from the precursor represented as $A$. This reaction follows an autocatalytic growth of nuclei leading to the formation of $B$ particles (see Fig. 2).

$$Nucleation: A \xrightarrow{k_1} B\ ;\quad n_0 = k_1[A], \tag{1}$$

$$Growth: A + B \xrightarrow{k_2} 2B;\quad G = k_2[A][B], \tag{2}$$

where, $n_0$ and $G$ are the nucleation and growth rates, $k_1$ and $k_2$ are the nucleation and growth kinetic constants and $[A]$ and $[B]$ are the molar concentrations of the precursor and silver in the nuclei/particles. The overall reaction rate can be expressed as shown:

$$-\frac{d[A]}{dt} = \frac{d[B]}{dt} = k_1[A] + k_2[A][B] \tag{3}$$

The F-W mechanism can be expressed though the integrated form (Eq. 4) which was also used to fit all experimental kinetic data sets [37]. It was considered that the concentration of material inside the particles changes with time $t$ as $[B]_t = [A]_0 - [A]_t$, where $[A]_0$ is the concentration of the precursor solution at $t = 0$ and $[A]_t$ the concentration of precursor at time $t$.

$$[B](t) = [A]_0 - \frac{\frac{k_1}{k_2} + [A]_0}{1 + \frac{k_1}{k_2[A]_0} * exp(k_1 + k_2[A]_0)\, t}. \tag{4}$$

For curve fitting procedures, the value of $[A]$ was estimated using Eq. (5), where $a$ is $B_t/B_\infty$ and $B_t$ and $B_\infty$ are the maximum absorbance at $t$ and $\infty$, respectively [43].

$$\frac{[A]_0 - [A]}{[A]} = \frac{a}{1-a} \tag{5}$$

The nucleation $k_1$ and growth $k_2$ constants were estimated using the linearized form below under assumptions that nucleation happens more slowly than growth (i.e. $k_1 \ll k_2[A]$) and $[A] < [A]_0$ [43].

$$ln\left(\frac{[A]_0 - [A]}{[A]}\right) = ln\left(\frac{k_1}{k_2[A_0]}\right) + k_2[A]_0 t, \tag{6}$$

By plotting, $ln\,(a/1 - a)$ versus $t$, a straight line with a slope and intercept were obtained as shown in Fig. 4B in Section 3.1. The values of rate constants were found by varying pH of reducing solution at two levels, pH 7 and pH 12. At each value of pH, three different concentrations of TC were probed: 1.91, 2.87 and 3.82 mM. This provided six different pairs of nucleation and growth constants to be used in the ML models.

## 2.2 Machine learning modelling

The machine learning analysis was performed using Python programming language with the Scikit-learn package [44]. Tree-based algorithms, including Decision Tree (DT), Random Forest (RF), and Extreme Gradient Boosting (XGBoost) were applied to test the prediction performance for the sizes of AgNPs. These methods have been extensively used in a wide range of engineering problems [45],[46],[47]. After implementation, all the proposed models were compared based on the accuracy of their predictions.

The DT algorithm used to guide the experiments was built based on 20 samples, each being repeated three times to obtain the averaged '(output quantity)' as model output. For fair comparison, 10 extra experiments (30 experiments in total, including three replications) were carried out either randomly or following the DT-guided design of experiments to enhance the performance of the predictive models. For each experiment, the average size was recorded. Each dataset was divided into five inputs and one predicted output. The training features (inputs) and their investigated range are summarized in Table 1.

$$Re = \frac{\rho u d}{\mu} \tag{7}$$

$$De = Re \sqrt{d/2R_c} \tag{8}$$

Eq. (7) and (8) define Reynolds, $Re$, and Dean, $De$, numbers respectively, where $d$ is the inner diameter of the microchannel, $R_c$ is the radius of helix curvature, $\rho$ is the fluid density, $u$ is the velocity of the fluid, and $\mu$ is the viscosity of the fluid.

Table 1: Input parameters and their investigated range.

| Feature | Range investigated |
| --- | --- |
| Nucleation constant ($min^{-1}$) | 0.0011 - 0.1 |
| Growth constant ($M^{-1} min^{-1}$) | 13.66-77.97 |
| Storage temperature (°C) | 0 - 20 |

| Dean number/Reynolds number | 0 - 0.41 |
| Reynolds number | 0.0849-16.96 |

The effect of the chosen input parameters on the size of AgNPs was investigated through the Pearson correlation [48] shown in Eq.(9) [49].

$$p_{XY} = \frac{\sum(x_i-\bar{x})(y_i-\bar{y})}{\sqrt{\sum(x_i-\bar{x})^2 \cdot \sum(y_i-\bar{y})^2}} \qquad (9)$$

In Eq. (9) $x_i$ is for the values of the independent variables, $\bar{x}$ is the mean of the values of x-variable, $y_i$ is for the values of the dependent variables and $\bar{y}$ is the mean of the values of y-variable.

**2.3 DT-guided design of experiments**

The iterative process of prediction and experiment design carried out is shown in Fig. 3, with the key focus being to perform experiments where the predictive model has the highest uncertainties. In particular, DT is chosen to guide the experiments thanks to its good interpretability and computational efficiency [50]. The first step in the proposed approach described previously was focused on obtaining the datasets required for the training of the algorithm, i.e., data collection based on experimental measurements. The second and third steps included data sorting or any other required data processing and the development of the most appropriate algorithm. For example, during these steps, the independent and dependent variables were identified, and a DT based algorithm was built. To ensure a good explainability of the DT model for the uncertainty evaluation, the maximum tree depth for this stage was set to be 4.

The fourth stage comprised carrying out additional experiments where the obtained DT algorithm has the highest mean squared error (MSE). The challenge of this step was to ensure that the developed DT model based on the designed experiments can improve the performance regarding different metrics. As shown in Fig. 3, once the experimental design is performed using the preliminary DT model, other ML classifiers such as RF and XGBoost can be applied to achieve a more accurate prediction based on the new experimental data.

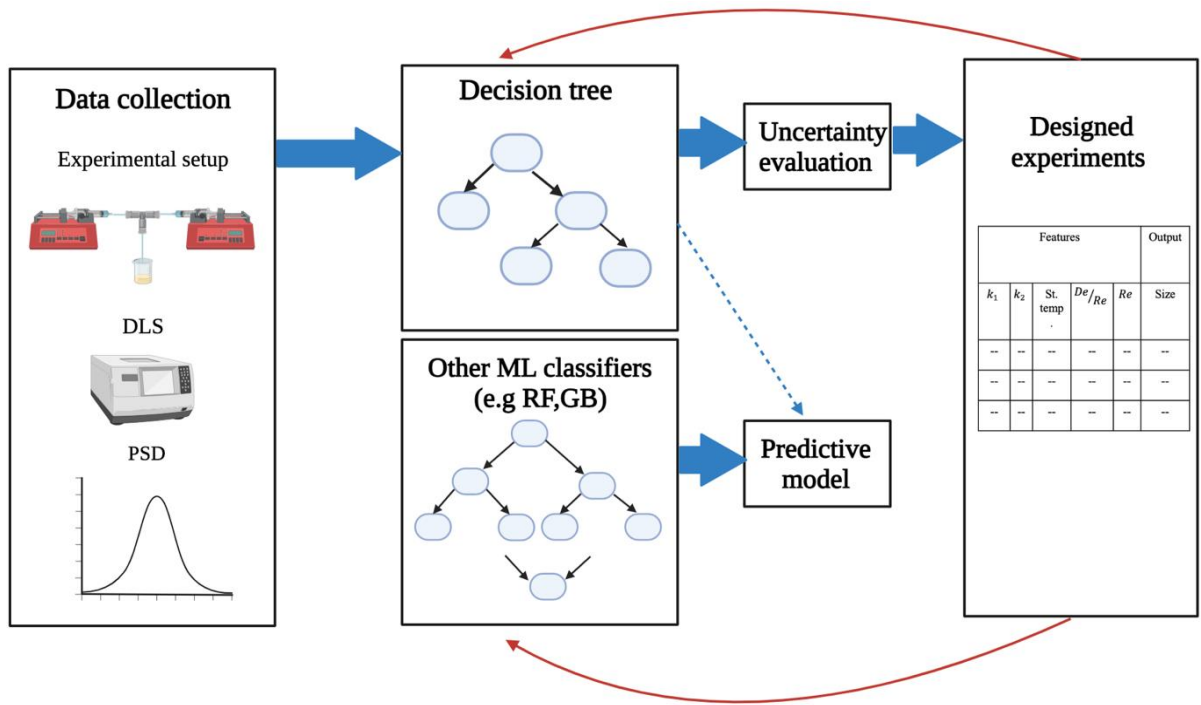

Figure 3: Workflow chart and description of methods used in the building of machine learning-guided design of experiment based on the decision tree method.

# 3. Results and discussion

## 3.1 Kinetic constants

The kinetic curves describing time dependence of concentration of silver included in nanoparticles (Fig.4A, 4B) were analysed by F-W mechanism [34] described in detail in Section 2.1. At pH 7, nucleation is rather slow, and the kinetic curves have a typical sigmoidal shape. The values of kinetic constants found from the linear fitting are given in Table 2. Using these values, it is easy to find that condition $k_1 \ll k_2[A]$ is valid for [A] >> 0.03 mM, which is only 3 % of initial concentration of SN, 0.92 mM. The obtained values of rate constants are validated by recalculating the full kinetic curve using Eq. (4) with these constants. Excellent agreement between the theoretical and experimental kinetic curves is shown in Fig. 4A (pH 7).

Multiple studies such as Anigol, Charantimath [5], Jebakumar Immanuel Edison and Sethuraman [51], and Dong, Ji [6] have shown that pH changes the kinetics in the formation of AgNPs, using *Capparis Moonii* fruit extract, Pod Extract of *Acacia nilotica* and citrate as reducing agents, respectively. At low pH, there is a slow reduction rate of the precursor while at high pH there is an improved reducing ability of the reducing agent that results in smaller size particles.

At pH 12, it was observed that the silver nuclei formation proceeded very fast, without a noticeable lag phase for all studied stabilizing agent concentrations, in agreement with literature data [35]. Only the right-hand side part of sigmoidal curve can thus be measured, which mostly accounts for particle growth. Although the linear fitting in Fig. 4D (pH 12) gives a reasonable fit, using values of kinetic constants provided in Table 2 shows that this linear fitting is valid only for [A] >> 0.5 mM for [TC] = 3.82 mM and [A] >> 1 mM for [TC] = 1.91 mM. Using these constants with the full F-W model, Eq. (4), does not provide as good agreement with experimental data as for the lower value of pH, but it is still acceptable as can be seen from Fig. 4D (pH 12). Note, non-linear fitting directly using Eq. (6) does not improve the fitting presented in Fig. 4B (pH 12). Thus, the values of nucleation and growth constants presented in Table 2 were used as the best possible fitting. For the investigated case, under alkaline pH, there is a full dissociation of citrate species [6] and dicarboxyacetone (DCA) [12], generated as a by-product of the decarboxylation of citrate during silver ions reduction, which makes the chemistry of the system more complex. Therefore, it is possible that several parallel processes are involved with different nucleation and growth constants making fitting of the kinetic curves much more challenging.

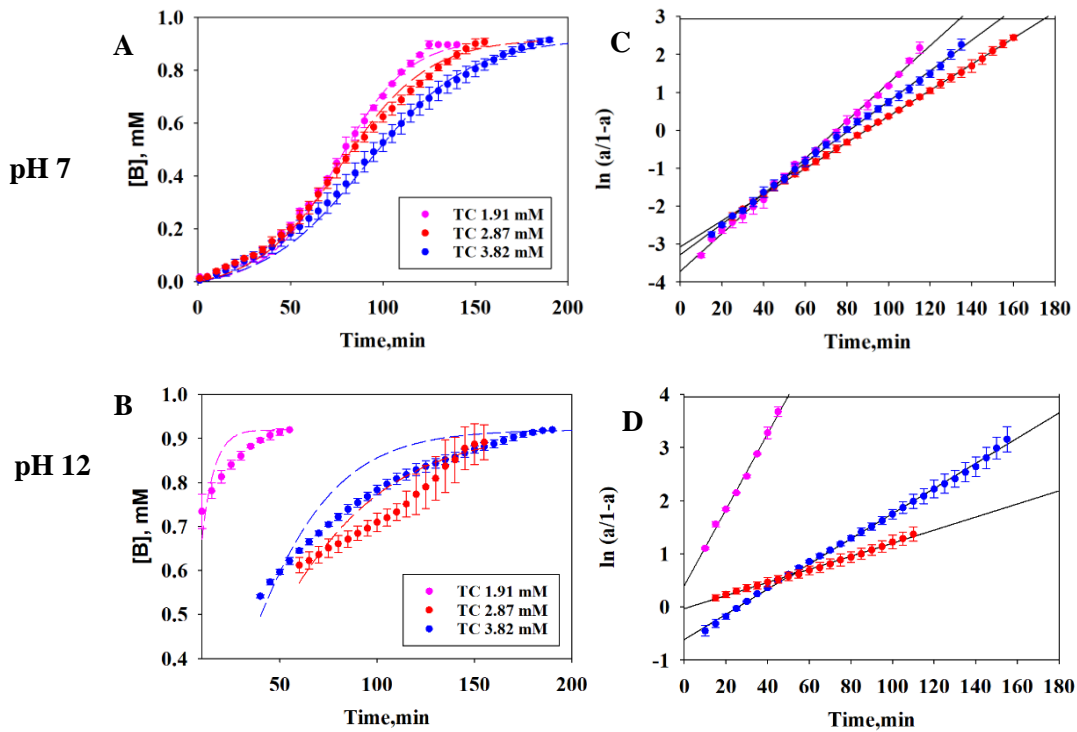

Figure 4: Application of the F-W mechanism for the analysis of AgNPs formation kinetics. Time evolution of [$B$] at 400 nm for A) pH 7 and B) pH 12 and different concentrations of the stabilizing agent: symbols represent experimental data; lines are fitting of full model Eq. (4) with kinetic constants found from linear fitting; Fitting of experimental data using linear form of Finke-Watzky model as shown in Eq. (6) for C) pH 7 and D) for pH 12.

Table 2: Nucleation and growth constants for varied pH and concentrations of trisodium citrate.

| pH | [TC] (mM) | Nucleation constant, $k_1$ ($min^{-1}$) (average) | Growth constant, $k_2$ ($M^{-1}min^{-1}$) (average) |
|---|---|---|---|
| 7 | 1.91 | 0.0011 | 55.51 |
| 7 | 2.87 | 0.0013 | 37.86 |
| 7 | 3.82 | 0.0015 | 44.81 |
| 12 | 1.91 | 0.1 | 77.97 |
| 12 | 2.87 | 0.012 | 13.66 |
| 12 | 3.82 | 0.0128 | 26.01 |

## 3.2 Pearson Correlation and the effect of input parameters

The results of Pearson correlation calculations are presented in Fig. 5 which shows that all input variables impact the particle size significantly with a correlation larger than 0.12 in absolute value. The estimation of the Pearson correlation value was obtained using all of the data generated in this study. The data used for this study can be found in the supporting information (SI). For example, it was observed that the variables including $k_1$, $k_2$ and $De/Re$ impact the particle size negatively, whilst the storage temperature and $Re$ impact the size of particles positively. The Pearson correlation calculations showed that $k_2$ is less important compared to $k_1$ for the prediction of the size of AgNPs. In this study, the effect of kinetic constants was investigated for only one set of chemicals, so that $k_1$ and $k_2$ cannot be considered as completely uncoupled. Therefore, it cannot be concluded that faster growth results in smaller particles, because for the chosen set of chemicals all growth constants are of the same order of magnitude, whereas nucleation constants vary over 2 orders of magnitude. Thus, the particle size is mostly defined by the nucleation constant. Similar studies, such as the works of Mansouri and Ghader [52] and Liu, Zhang [53] showed that size of particles is a result of a significant effect of $k_1$ instead of $k_2$.

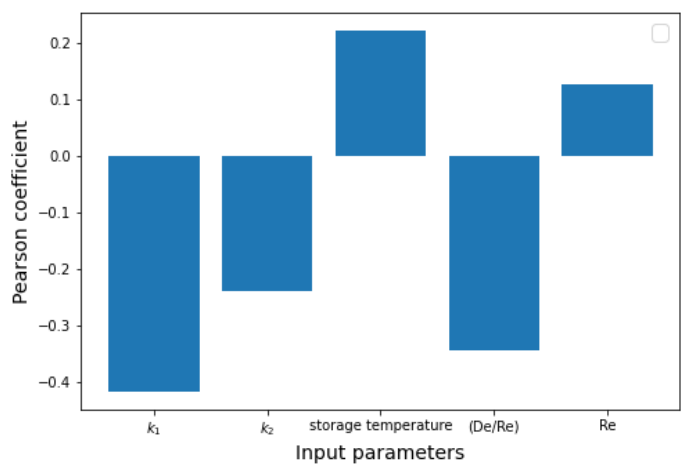

Figure 5: Pearson correlation between input variables and the size of AgNPs

The samples of AgNPs were kept at 20 °C and 0 °C and Fig. 5 shows that the temperature has a positive effect on the particle size, i.e., that smaller particles were observed at low storage temperatures. The full set of data on the particles size depending on process parameters, including storage temperature is provided in SI. Measurements of the silver nitrate concentration for different flow rates of reagents show that the reduction reaction is completed before collection. For example, at $Re = 5.66$ with a 2 m outlet tube, the residual concentration of silver nitrate was around $0.017 \pm 0.00053$ mM (less than 2 % of initial concentration) whilst at highest flow rate used, $Re = 16.96$ and 2 m length, the residual concentration of silver nitrate was 0mM ($\sim$ 3% of initial concentration). The small residual concentrations of silver nitrate showed that the reaction was practically complete for all the investigated values of *Re* including the highest one where residence time was smaller. Izak-Nau, Huk [54] and Peng, Krauss [55] have described in detail the ageing of AgNPs samples stored at room temperature, proposing that growth of the particle size is due to agglomeration or oxidation of silver.

When helical coils are employed, improved mixing due to the formation of Dean vortices leads to smaller particles, and this effect is one of the largest (Fig. 5). It can be assumed that a decrease in particle size due to faster mixing is the result of more intensive nuclei formation at the initial stage of reaction. Of course, faster mixing results also in faster particle growth, but comparing the relative effect of nucleation and growth constant in Fig. 5, it can be suggested that effect of faster nucleation is larger than the effect of faster growth. The effect of the Reynolds number is positive, yet the least significant.

### 3.3 Prediction results with DT-guided experiments

The performance of the developed models (DT, RF, XGBoost) was described through MSE, MAE and $R^2$. In the following Eqs( 10)-(12), $n$ is the number of data within training and testing data sets, $X_i$ is the predicted $i^{th}$ value, $Y_i$ is the actual $i^{th}$ value and $\hat{Y}$ is the mean of true values [56].

$$MSE = \frac{1}{n}\sum_{i=1}^{n}(X_i - Y_i)^2 \qquad (10)$$

$$MAE = \frac{1}{n}\sum_{i=1}^{n}|X_i - Y_i| \qquad (11)$$

$$R^2 = 1 - \frac{\sum_{i=1}^{n}(X_i-Y_i)^2}{\sum_{i=1}^{n}(\hat{Y}-Y_i)^2} \qquad (12)$$

The decision tree shown in Fig.6 examined the synthesis of AgNPs based on the combination of five conditions. The decision tree shows different routes from up to down for preparing AgNPs. Each box presents the value of the parameter, the estimated size, the MSE and the sample sizes. The tree showed that there is a high MSE (Eq. (10)) in some synthesis routes e.g., when the growth constant, $k_2$ was smaller than 32, the storage temperature was higher than 10 °C, and the values of $De/Re$ and $Re$ were smaller and higher than 0.2 and 0.47 respectively. To improve the performance of the DT developed, synthesis routes were designed under the conditions with high MSE (orange routes on DT). These designed experiments were used not only to improve the performance of the developed DT model but also of other models such us RF and XGBoost. Additionally, the designed synthesis routes (set of parameters identified in the DT with high MSE) were applied to confirm the validity of this approach.

The closer the value of $R^2$ is to 1 and the values of MAE and MSE are to zero, the better is the fit of the developed models. The statistical characteristics of the three appropriate models for the original (20 initial conditions used to build the DT model), random (10 random extra experiment conditions) and designed extra data sets (10 extra designed experiment conditions where high uncertainty was found) are given in detail in Table 3.

These results are obtained on the unseen test data. In particular, the designed data sets are obtained by uncertainty analysis (on the training data) of preliminary DT predictions as illustrated in Fig. 6. As shown, the values of $R^2$ are always increasing with the designed experiments for all the models while MAE and MSE are decreasing. These results confirmed that the designed experiments based on DT approach can improve the final predictive models with less cost and time.

As shown in Fig. 7 and Table 3, the computed predictive models, including DT, RF, and XGBoost, which were assisted by DT-based designed of experiments, outperform considerably the ones trained on random extra experiments. These results also confirmed that the proposed DT-based on designed experiments can significantly enhance the performance of other machine learning algorithms with a considerable reduction (-26% for XGBoost and -38% for RF) of MSE compared to the original model. Note, this paper consists of a 'proof of concept' of the proposed DT-guided design of experiments and more experimental data should be collected to obtain more reliable predictions.

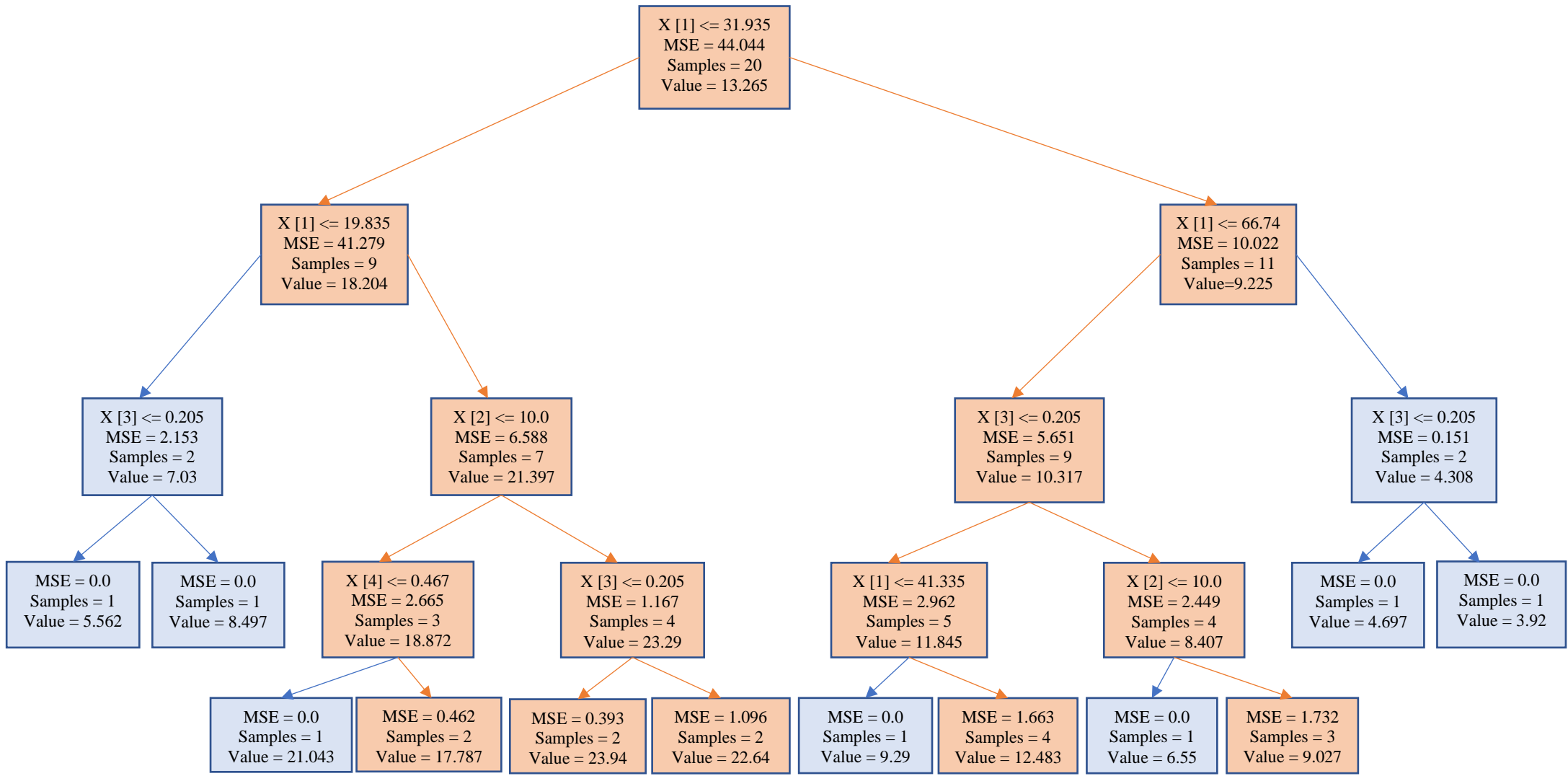

Figure 6: Decision tree which examines the size of synthesized AgNPs as a function of five parameters where $x_0, x_1, x_2, x_3, x_4$, are nucleation constant $k_1$, growth constant $k_2$, storage temperature, Dean number/Reynolds number ($De/Re$) and Reynolds number ($Re$) respectively. Orange boxes are showing where there is a high MSE.

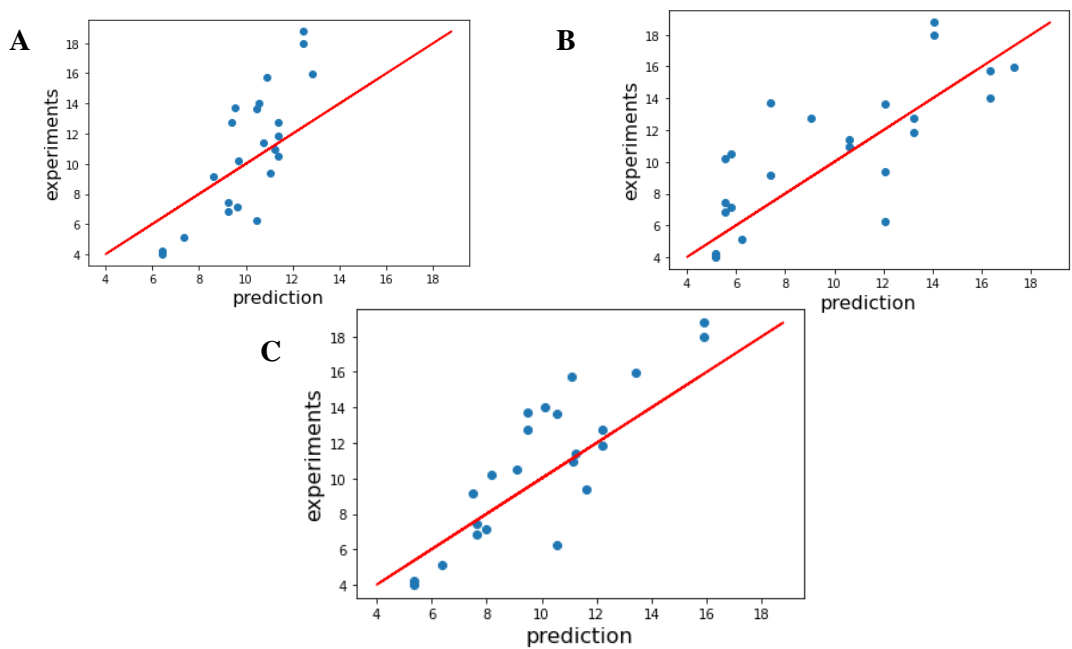

Figure 7: Predicted versus real measurements for the size of AgNPs using DT. A) Original training data (20 initial conditions, 60 training samples in total including replications), B) Random extra experiments (10 experiments, 30 validation samples in total including replications) and C) Designed extra experiments (10 experiments, 30 validation samples in total including replications). The conditions that have been used for training, random extra experiments and designed extra experiments are described in detail in SI. The average size of AgNPs on X, Y axes is given in nm.

Table 3: Decision tree (DT), Extreme Gradient boosting (XGBoost) and Random Forest (RF) results for original, random and designed experiments.

A. Decision Tree

|  | MSE | MAE | $R^2$ |
| --- | --- | --- | --- |
| **Original** | 9.43 | 2.46 | 0.45 |
| **Random** | 8.65 | 2.24 | 0.49 |
| **Designed** | 6.41 | 2.01 | 0.62 |

B. Gradient Boosting

|  | MSE | MAE | $R^2$ |
| --- | --- | --- | --- |
| **Original** | 8.88 | 2.27 | 0.47 |
| **Random** | 7.26 | 2.01 | 0.57 |
| **Designed** | 6.56 | 2.16 | 0.61 |

C. Random Forest

|          | MSE  | MAE  | $R^2$ |
|----------|------|------|-------|
| **Original** | 9.42 | 2.52 | 0.44  |
| **Random**   | 6.45 | 1.93 | 0.62  |
| **Designed** | 5.82 | 1.96 | 0.66  |

**4.Conclusions**

A machine learning-guided design of experiments based on the decision tree method for the size of AgNPs synthesized in a continuous flow T-junction device was implemented. The parameters investigated were nucleation and growth constants derived from an independent set of experiments using F-W mechanism, as well as storage temperature and hydrodynamic parameters including Reynolds and Dean number.

The decision tree based on designed experiments is an effective and low-cost strategy to improve a system in a much more directed manner. The obtained model showed the areas of interest for further experiments (regions with high MSE) and numerical results showed that well designed synthesis routes can improve significantly the performance not only of DT itself but also of other regression methods such as RF and XGBoost compared to randomly extra synthesis routes. In particular, considerable reduction of MSE compared to the original model, 26% for XGBoost and 38% for RF, was found.

Additionally, the obtained model can be developed further in the future by data-assimilation from numerical simulations and additional experiments, including experiments with other chemicals and other device types.

**Declaration of Competing Interest**

The authors declare that they have no known competing financial interests or personal relationships that could have appeared to influence the work reported in this paper.

**Acknowledgements**

The authors gratefully acknowledge support from the Engineering & Physical Sciences Research Council, UK, through the PREMIERE Programme Grant (EP/T000414/1). KN is funded by a PhD scholarship from the School of Chemical Engineering, University of Birmingham.

**Supporting information**

Table S1: Particles sizes of AgNPs at different conditions. The sizes were measured as a function of nucleation and growth constants, storage temperature, Reynolds number and Dean number/Reynolds number.

| Nucleation constant (min$^{-1}$) | Growth constant (M$^{-1}$min$^{-1}$) | Storage temperature (°C) | Dean number / Reynolds number | Reynolds number | AgNPs size (1) | AgNPs size (2) | AgNPs size (3) |
|---|---|---|---|---|---|---|---|
| \multicolumn{8}{c}{Original experiments} ||||||||
| 0.0128 | 26.01 | 20 | 0.41 | 3.395 | 19.05 | 22.08 | 23.65 |
| 0.0015 | 44.81 | 0 | 0 | 3.395 | 11.1 | 13.03 | 12 |
| 0.0015 | 44.81 | 20 | 0.41 | 3.395 | 8.53 | 12.2 | 11.16 |
| 0.0128 | 26.01 | 0 | 0 | 3.395 | 17.62 | 16.71 | 16.99 |
| 0.0015 | 44.81 | 0 | 0.41 | 3.395 | 5.68 | 7.35 | 6.62 |
| 0.0128 | 26.01 | 20 | 0 | 3.395 | 23.38 | 25.52 | 24.8 |
| 0.0128 | 26.01 | 0 | 0.41 | 0.0849 | 19.83 | 19.81 | 23.49 |
| 0.0015 | 44.81 | 20 | 0 | 0.0849 | 12.64 | 13.54 | 13.54 |
| 0.1 | 77.97 | 0 | 0.41 | 3.395 | 4.39 | 3.58 | 3.79 |
| 0.0013 | 37.86 | 20 | 0 | 3.395 | 9.68 | 8.77 | 9.42 |
| 0.0128 | 26.01 | 0 | 0 | 0.849 | 19.77 | 17.39 | 18.24 |
| 0.0011 | 55.51 | 0 | 0 | 3.395 | 15.52 | 15.24 | 11.34 |
| 0.1 | 77.97 | 20 | 0 | 3.395 | 5.08 | 4.26 | 4.75 |
| 0.0015 | 44.81 | 20 | 0.41 | 0.849 | 6.73 | 7.54 | 7.95 |
| 0.012 | 13.66 | 20 | 0.41 | 3.395 | 9.53 | 7.28 | 8.68 |
| 0.012 | 13.66 | 0 | 0 | 3.395 | 5.3 | 6.097 | 5.29 |
| 0.0011 | 55.51 | 20 | 0.41 | 3.395 | 8.1 | 10.2 | 8.83 |
| 0.0015 | 44.81 | 0 | 0 | 0.0849 | 11.41 | 10.37 | 10.07 |
| 0.0128 | 26.01 | 20 | 0.41 | 0.0849 | 23.93 | 23.93 | 23.2 |
| 0.0128 | 26.01 | 20 | 0 | 0.849 | 21.52 | 22.95 | 25.47 |

| Designed experiments based on decision tree | | | | | | | |
|---|---|---|---|---|---|---|---|
| 0.0011 | 55.51 | 20 | 0.41 | 0.0849 | 8.932 | 10.141 | 11.35 |
| 0.0011 | 55.51 | 0 | 0 | 0.849 | 11.63 | 12.9 | 15.41 |
| 0.0128 | 26.01 | 0 | 0.41 | 0.849 | 15.7 | 14.28 | 17.1 |
| 0.0013 | 37.86 | 0 | 0 | 0.0849 | 12.19 | 11.34 | 10.91 |
| 0.0015 | 44.81 | 0 | 0 | 0.849 | 11.42 | 11.92 | 10.58 |
| 0.0128 | 26.01 | 20 | 0 | 0.0849 | 20.78 | 24.4 | 23.01 |
| 0.0013 | 37.86 | 20 | 0.41 | 0.0849 | 7.161 | 6.45 | 8.69 |
| 0.0011 | 55.51 | 20 | 0.41 | 0.849 | 7.87 | 7.074 | 6.27 |
| 0.0013 | 37.86 | 0 | 0 | 0.849 | 10.1 | 8.78 | 9.38 |
| 0.0011 | 55.51 | 20 | 0 | 3.395 | 18.52 | 16.47 | 18.2 |
| Random experiments | | | | | | | |
| 0.1 | 77.97 | 0 | 0.41 | 0.849 | 4.55 | 3.54 | 4.547 |
| 0.0013 | 37.86 | 0 | 0 | 3.395 | 6.414 | 6 | 6.4 |
| 0.0015 | 44.81 | 20 | 0 | 0.849 | 12.95 | 13.65 | 11.67 |
| 0.0013 | 37.86 | 20 | 0.41 | 0.849 | 9.034 | 8.58 | 9.88 |
| 0.1 | 77.97 | 0 | 0.41 | 1.698 | 3.98 | 3.68 | 4.38 |
| 0.012 | 13.66 | 0 | 0.41 | 1.698 | 6.996 | 7.38 | 6.996 |
| 0.1 | 77.97 | 20 | 0.41 | 1.698 | 5.22 | 5.27 | 4.94 |
| 0.0015 | 44.81 | 20 | 0 | 1.698 | 13.98 | 11.7 | 9.878 |
| 0.0011 | 55.51 | 0 | 0 | 1.698 | 16.2 | 15.65 | 16.11 |
| 0.0011 | 55.51 | 20 | 0 | 16.96 | 17.48 | 19.86 | 19.01 |
| 0.0011 | 55.51 | 20 | 0 | 5.662 | 15.45 | 20.416 | 18.1 |
| 0.0011 | 55.51 | 20 | 0.32 | 0.0849 | 14.05 | 13.87 | 13.25 |
| 0.0011 | 55.51 | 20 | 0.32 | 3.395 | 12.14 | 13.4 | 12.614 |
| 0.012 | 13.66 | 20 | 0.41 | 0.0849 | 9.91 | 10.29 | 11.31 |
| 0.012 | 13.66 | 0 | 0 | 0.849 | 7.45 | 8.7 | 6.1 |
| 0.012 | 13.66 | 20 | 0 | 3.395 | 9.24 | 11.6 | 9.7 |
| 0.012 | 13.66 | 0 | 0 | 1.698 | 7.73 | 6.45 | 6.28 |
| 0.0015 | 44.81 | 0 | 0 | 0.424 | 10.96 | 10.414 | 11.39 |
| 0.0015 | 44.81 | 0 | 0 | 1.698 | 12.48 | 10.78 | 10.95 |
| 0.0015 | 44.81 | 0 | 0 | 2.546 | 9.77 | 9.13 | 9.19 |
| 0.0013 | 37.86 | 0 | 0 | 0.424 | 16.84 | 14.7 | 15.67 |
| 0.0013 | 37.86 | 0 | 0 | 2.546 | 13.73 | 13.57 | 13.7 |
| 0.0013 | 37.86 | 0 | 0 | 1.698 | 15.6 | 13.14 | 13.36 |

**5.References**